\documentclass[twocolumn, tighten, times]{aastex631}

\begin{document}

\title{The JWST Early Release Observations}

\correspondingauthor{Klaus M. Pontoppidan}
\email{pontoppi@stsci.edu}

\author[0000-0001-7552-1562]{Klaus M. Pontoppidan}
\affiliation{Space Telescope Science Institute, 3700 San Martin
  Drive, Baltimore, MD, 21218, USA}
\author{Jaclyn Barrientes}
\affiliation{Space Telescope Science Institute, 3700 San Martin
  Drive, Baltimore, MD, 21218, USA}
\author{Claire Blome}
\affiliation{Space Telescope Science Institute, 3700 San Martin
  Drive, Baltimore, MD, 21218, USA}
\author{Hannah Braun}
\affiliation{Space Telescope Science Institute, 3700 San Martin
  Drive, Baltimore, MD, 21218, USA}
\author{Matthew Brown}
\affiliation{Space Telescope Science Institute, 3700 San Martin
  Drive, Baltimore, MD, 21218, USA}
\author{Margaret Carruthers} 
\affiliation{Space Telescope Science Institute, 3700 San Martin
  Drive, Baltimore, MD, 21218, USA}
\author[0000-0001-7410-7669]{Dan Coe}
\affiliation{Space Telescope Science Institute, 3700 San Martin
  Drive, Baltimore, MD, 21218, USA}
\affiliation{Association of Universities for Research in Astronomy (AURA), Inc.
for the European Space Agency (ESA)}
\author[0000-0002-9142-9755]{Joseph DePasquale} 
\affiliation{Space Telescope Science Institute, 3700 San Martin
  Drive, Baltimore, MD, 21218, USA}
\author[0000-0001-9513-1449]{N\'estor Espinoza}
\affiliation{Space Telescope Science Institute, 3700 San Martin
  Drive, Baltimore, MD, 21218, USA}
\affiliation{Department of Physics \& Astronomy, Johns Hopkins University, Baltimore, MD 21218, USA}
\author{Macarena Garcia Marin}
\affiliation{Space Telescope Science Institute, 3700 San Martin
  Drive, Baltimore, MD, 21218, USA}
\affiliation{European Space Agency, Space Telescope Science Institute, Baltimore, Maryland, USA}  
\author[0000-0001-5340-6774]{Karl D.\ Gordon}
\affiliation{Space Telescope Science Institute, 3700 San Martin
  Drive, Baltimore, MD, 21218, USA}
\author[0000-0002-6586-4446 ]{Alaina Henry}
\affiliation{Space Telescope Science Institute, 3700 San Martin
  Drive, Baltimore, MD, 21218, USA}
\author{Leah Hustak}
\affiliation{Space Telescope Science Institute, 3700 San Martin
  Drive, Baltimore, MD, 21218, USA}
\author{Andi James}
\affiliation{Space Telescope Science Institute, 3700 San Martin
  Drive, Baltimore, MD, 21218, USA}
\author{Ann Jenkins}
\affiliation{Space Telescope Science Institute, 3700 San Martin
  Drive, Baltimore, MD, 21218, USA}
\author[0000-0002-6610-2048]{Anton M. Koekemoer}
\affiliation{Space Telescope Science Institute, 3700 San Martin
  Drive, Baltimore, MD, 21218, USA}
\author[0000-0002-5907-3330]{Stephanie LaMassa}
\affiliation{Space Telescope Science Institute, 3700 San Martin
  Drive, Baltimore, MD, 21218, USA}
\author[0000-0002-9402-186X]{David Law}
\affiliation{Space Telescope Science Institute, 3700 San Martin
  Drive, Baltimore, MD, 21218, USA}
\author{Alexandra Lockwood}
\affiliation{Space Telescope Science Institute, 3700 San Martin
  Drive, Baltimore, MD, 21218, USA}
\author[0000-0001-9504-8426]{Amaya Moro-Martin}
\affiliation{Space Telescope Science Institute, 3700 San Martin
  Drive, Baltimore, MD, 21218, USA}
\author[0000-0001-7106-4683]{Susan E. Mullally}
\affiliation{Space Telescope Science Institute, 3700 San Martin
  Drive, Baltimore, MD, 21218, USA}
\author[0000-0002-5359-5357]{Alyssa Pagan}
\affiliation{Space Telescope Science Institute, 3700 San Martin
  Drive, Baltimore, MD, 21218, USA}
\author{Dani Player}
\affiliation{Space Telescope Science Institute, 3700 San Martin
  Drive, Baltimore, MD, 21218, USA}
\author[0000-0001-7617-5665]{Charles Proffitt}
\affiliation{Space Telescope Science Institute, 3700 San Martin
  Drive, Baltimore, MD, 21218, USA}
\author{Christine Pulliam}
\affiliation{Space Telescope Science Institute, 3700 San Martin
  Drive, Baltimore, MD, 21218, USA}
\author{Leah Ramsay}
\affiliation{Space Telescope Science Institute, 3700 San Martin
  Drive, Baltimore, MD, 21218, USA}
\author[0000-0002-5269-6527]{Swara Ravindranath}
\affiliation{Space Telescope Science Institute, 3700 San Martin
  Drive, Baltimore, MD, 21218, USA}
\author{Neill Reid}
\affiliation{Space Telescope Science Institute, 3700 San Martin
  Drive, Baltimore, MD, 21218, USA}
\author[0000-0002-9573-3199]{Massimo Robberto}
\affiliation{Space Telescope Science Institute, 3700 San Martin
  Drive, Baltimore, MD, 21218, USA}
\author[0000-0003-2954-7643]{Elena Sabbi}
\affiliation{Space Telescope Science Institute, 3700 San Martin
  Drive, Baltimore, MD, 21218, USA}
\author[0000-0001-7130-2880]{Leonardo Ubeda}
\affiliation{Space Telescope Science Institute, 3700 San Martin
  Drive, Baltimore, MD, 21218, USA}
\author[0000-0003-4849-9536]{Michael Balogh}
\affiliation{Department of Physics and Astronomy, University of Waterloo, Waterloo, ON N2L 3G1, Canada}
\author{Kathryn Flanagan}
\affiliation{Space Telescope Science Institute, 3700 San Martin
  Drive, Baltimore, MD, 21218, USA}
\author[0000-0003-2098-9568]{Jonathan Gardner}
\affiliation{Astrophysics Science Division, NASA Goddard Space Flight Center, 8800 Greenbelt Rd, Greenbelt, MD 20771, USA}
\author{Hashima Hasan}
\affiliation{NASA Headquarters, 300 E St. SW, Washington, D.C. 20546, USA}
\author[0000-0001-5931-9305]{Bonnie Meinke}
\affiliation{National Ecological Observatory Network, 1685 38th St., Boulder, CO 80301, USA}
\author{Antonella Nota}
\affiliation{Space Telescope Science Institute, 3700 San Martin
  Drive, Baltimore, MD, 21218, USA}



\begin{abstract}
The James Webb Space Telescope (JWST) Early Release Observations (EROs) is a set of public outreach products created to mark the end of commissioning and the beginning of science operations for JWST. Colloquially known as the ``Webb First Images and Spectra", these products were intended to demonstrate to the worldwide public that JWST is ready for science, and is capable of producing spectacular results. The package was released on July 12, 2022, and included images and spectra of the galaxy cluster SMACS~J0723.3-7327 and distant lensed galaxies, the interacting galaxy group Stephan's Quintet, NGC 3324 in the Carina star-forming complex, the Southern Ring planetary nebula NGC 3132, and the transiting hot Jupiter WASP 96b. This paper describes the ERO technical design, observations, and scientific processing of data underlying the colorful outreach products.
\end{abstract}



\section{Purpose and design} \label{sec:intro}
The JWST Early Release Observations (EROs) were designed to demonstrate that the observatory is ready for science, and is capable of producing spectacular results. This follows the precedents set by the Chandra and Spitzer space telescopes, which produced similar EROs after their launch and successful commissioning, and by the Hubble Space Telescope subsequent to each servicing mission. The observations were to be used to produce a colorful set of images and spectra for a public press release. However, given the great leap in sensitivity and resolution of JWST compared to the previous generations of infrared telescopes, the ERO data are also available to be analyzed for scientific purposes. 

A significant subset of the EROs were used to create content for the ``Webb First Images and Spectra" press release, with one image (``Webb's First Deep Field", SMACS J0723.3-7327) presented by President Joseph Biden on July 11, 2022, and the remainder released at a NASA press briefing the next day, on July 12, 2022. The press release was one of the largest public science events ever, resulting in $\sim$ 26,000 news articles and $\sim$120 billion impressions over the course of a few days. The metrics were extracted using the Meltwater media monitoring service for the period spanning July 12-16th. The number of impressions are based on total visitors to the websites on which the articles appeared. 

The science data underlying the First Images and Spectra were publicly released on the Mikulski Archive for Space Telescopes (MAST) on July 13th, 2022. The observations highlight key science areas where JWST is most likely to make transformational contributions to astrophysics: Early galaxies, interacting galaxies, stellar birth and death, and other worlds (i.e., exoplanets). 

This paper presents the JWST First Images and Spectra data, describes their observational design, processing, and visualization steps carried out to support the press event. We do not present any scientific analysis of the data set, and leave this to the astronomical community. All the public outreach products resulting from this set of observations can be found on \url{webbtelescope.org}. 

\section {ERO Target Selection and Production Support }

The formal responsibility for selecting the JWST ERO targets rested with the JWST Program Officer at NASA Headquarters, based on advice from an ERO Selection Committee including representatives selected by each agency\footnote{Constituted in 2016, the ERO Selection Committee members included Michael Balogh (University of Waterloo, for CSA), Jonathan Gardner (NASA, GSFC), Hashima Hasan (NASA HQ), and Antonella Nota (ESA), with support by STScI staff as Executive Secretary (initially Elena Sabbi, latterly Amaya Moro-Martin) and a Coordination Scientist (initially Kathryn Flanagan, followed by Bonnie Meinke, and then Neill Reid). The JWST Program Officer at HQ is Eric Smith.}. The ERO committee identified potential targets from a wide set of sources, including the JWST Science Working Group in May 2017. Consideration was given to seeking direct public input, but it was unclear how such input could be prioritized in an unbiased manner. Instead, the committee polled the members of the American Astronomical Society, soliciting target selections in late 2017\footnote{https://aas.org/posts/letter/2017/10/input-welcome-jwsts-early-release-observations}.  

Based on the input received, the ERO Committee identified a superset of targets that might be suitable for JWST’s first public science observations. Those targets were evaluated based on existing data, particularly Hubble and Spitzer Space Telescope color images, and on their likely infrared characteristics, technical feasibility, as well as their relevance to JWST’s four science themes. The final prioritized list was delivered in early 2020 and included some 70 targets distributed across the full sky, providing contingency to cover a wide range of potential launch dates. 

Working from that list, the ERO Implementation Core Team\footnote{ERO Implementation Core Team members – Klaus Pontoppidan, Alex Lockwood, Amaya Moro-Martin, Hannah Braun, Neill Reid} identified the highest priority targets whose availability was consistent with the December 25, 2021 launch. The final targets span JWST’s science themes and include representative observations from all major modes of JWST’s four science instruments. In summer 2021, members of STScI’s JWST instrument branches were recruited to assist in producing and verifying the Astronomer Proposal Tool observing files, and to provide crucial support for data reduction. Finally, in early 2022, representatives from STScI’s Office of Public Outreach joined the ERO Production Team to contribute expertise in graphics design, science writing, and news production.  

The JWST First Images and Spectra were created by Space Telescope Science Institute (STScI) staff between June 3 and July 10, 2022, from the first observation to the final delivery to NASA. In total, more than 30 people were involved in the production team, supported by the full commissioning and operations system for the JWST observatory. The data sets described here will ultimately be made available in MAST as high-level science products (HLSPs), once all necessary calibration files are available.

\section{Observations}

The JWST ERO observations were planned and executed as part of the JWST commissioning program. Programs included in the final release were PID 2731, 2732, 2733, 2734, and 2736 (PI: Pontoppidan). The data are available at the Mikulski Archive for Space Telescopes \footnote{\dataset[10.17909/kjms-sq75]{\doi{10.17909/kjms-sq75}}}.

In general, planned observations were modeled using the JWST Exposure Time Calculator \citep{Pontoppidan16}, using archival imaging from Spitzer IRAC \citep[3.6-8.0\,$\mu$m;][]{Fazio04} and \citep[24\,$\mu$m;][]{Rieke04} MIPS. 5-28\,$\mu$m spectra from the Spitzer Infrared Spectrometer \citep{Houck04}, where available, were used to inform filter choices, supplemented by ground-based 1-2\,$\mu$m spectroscopy from various sources in the literature. 

\subsection{SMACS J0723.3-7327}

SMACS J0723.3-7327 is a massive galaxy cluster (z=0.388) known to act as a strong gravitational lens, magnifying the background field of distant, high-redshift galaxies. This cluster was first discovered as part of the southern extension of the Massive Cluster Survey  \cite[MACS:][]{Ebeling01}, and its properties were described by \cite{Repp18}. 
It was first observed by Hubble in programs 11103, 12166 and 12884 (PI: Ebeling), then later as part of the {\it The Reionization Lensing Cluster Survey (RELICS)} survey \citep{Coe19}.
It was selected for this ERO program for its bright central elliptical galaxy, prominently lensed arcs, and high ecliptic latitude at $-$80.2\arcdeg\ yielding a low Zodiacal background. 

\begin{figure*}[ht!]
    \centering
    \includegraphics[width=18cm]{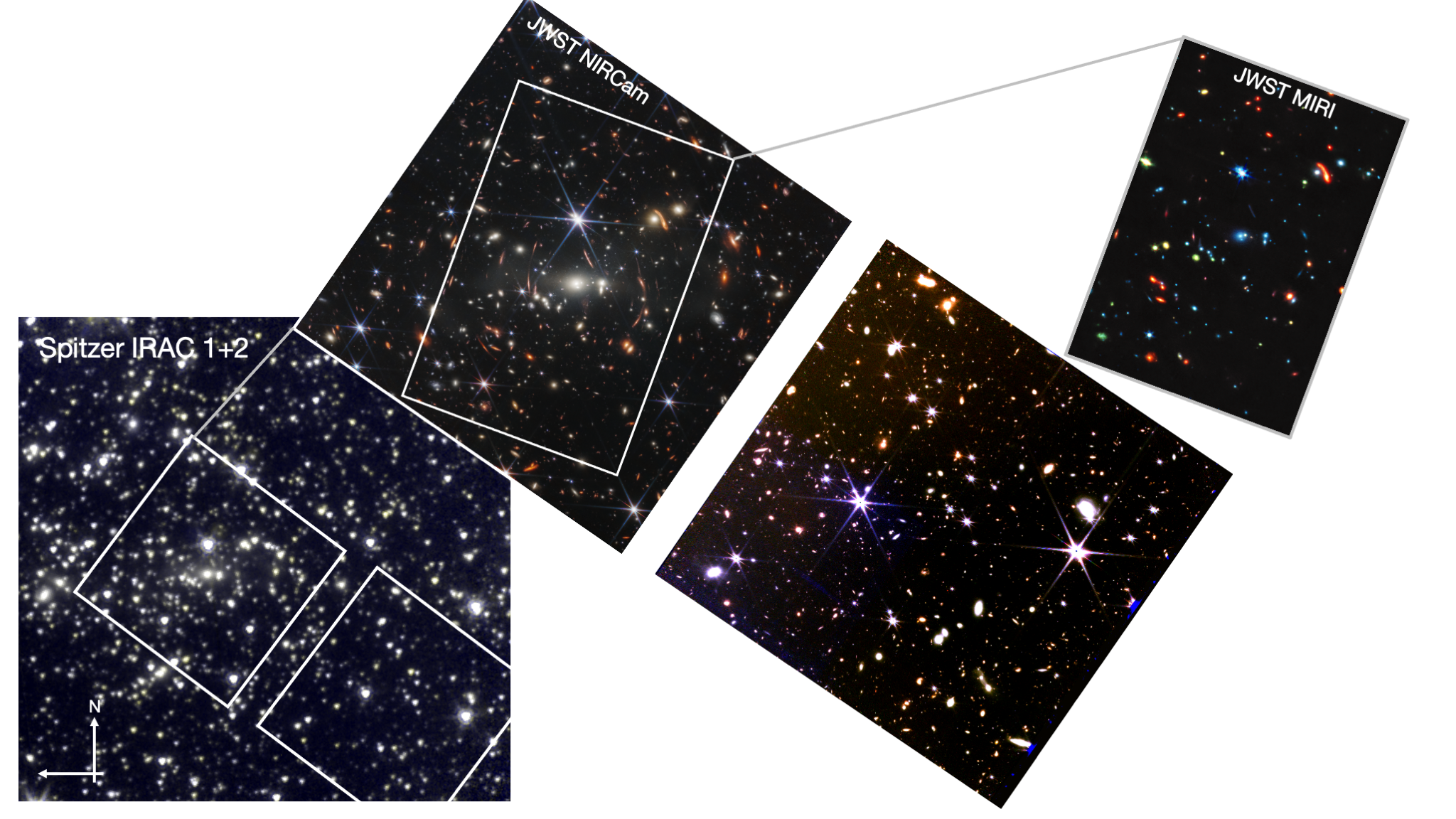}
    \caption{Field of view in SMACS J0723.3-7327 covered by NIRCam and MIRI. Full-resolution JWST images are available on webbtelescope.org.}
    \label{fig:smacs_field}
\end{figure*}

The ERO multi-mode observations of SMACS J0723.3-7327 include NIRCam and MIRI imaging of the cluster and surrounding field (Fig.~\ref{fig:smacs_field}), supporting NIRSpec Multi-Object Spectroscopy, and NIRISS Wide-Field Slitless Spectroscopy. The aim of the observations was to demonstrate the ability of JWST to rapidly image high-redshift galaxies, at a depth rivaling the most sensitive Hubble deep fields. The gravitationally lensed arcs both increase the richness of the background field, as well as offer a science narrative related to the use of gravitational lensing to magnify and enhance distant galaxies. The multi-object spectroscopy was intended to demonstrate emission-line signatures of star-forming galaxies, and illustrate to the public how redshifts (look-back time) can be measured to high precision by JWST. The NIRCam imaging was also used to create a catalog from which the NIRSpec MSA configuration could be constructed, following a workflow similar to that which future science programs would use. 

\begin{deluxetable*}{ccccc}[ht!]
\tablecaption{Observation parameters for SMACS J0723.3-7327 / PID 2736}
\label{smacs_details}
\tablehead{
\colhead{Instrument} & \colhead{Mode} & \colhead{Filter/Disperser} & \colhead{Exposure Time} & \colhead{Observing date}\\
&&& Seconds per Filter/Disperser & 
} 
\startdata
NIRCam & Imaging & F090W/F150W/F200W/F277W/F356W/F444W & 7,537 & 2022 Jun 7\\
MIRI & Imaging & F770W/F1000W/F1500W & 5,578 & 2022 Jun 14 \\
MIRI & Imaging & F1800W & 5,633 & 2022 Jun 14 \\
NIRISS & WFSS & GR150R+F115W/GR150R+F200W & 2,835 & 2022 Jun 17 \\
NIRISS & WFSS & GR150C+F115W/GR150C+F200W & 2,835 & 2022 Jun 17 \\
NIRSpec & MSA & G235M/G395M & 17,682 & 2022 Jun 30 \\
\enddata
\end{deluxetable*}

\subsubsection{NIRCam imaging} 
The NIRCam image centers module B on the cluster, and module A on an adjacent off-field. While the ERO product featured only the cluster module, the off-field module represents an additional, serendipitous deep field. We obtained NIRCam images through 6 broad-band NIRCam filters, spanning F090W through F444W (see Table \ref{smacs_details}). The exposure time was designed to achieve a point source sensitivity of AB$\sim\,$29.8\,mag, as estimated using the JWST Exposure Time Calculator (5$\sigma$), aimed at matching the depth of the Hubble Ultra Deep Field as observed with WFC3/IR \citep[HUDF12/XDF:][]{Ellis13, Koekemoer13, Illingworth13}, and surpassing by at least a factor of 10 the depth of Spitzer IRAC channel 1+2 imaging for similar fields. In practice, due to improved NIRCam sensitivity in these filters by up to 30\%, as uncovered during commissioning \citep{Rigby22}, combined with the lower Zodiacal background of this field compared with the HUDF, the ultimate depth of the SMACS J0723.3-7327 field may be somewhat deeper, perhaps reaching or surpassing $AB\sim 30$\,mag over much of the field. 

We used the INTRAMODULEX dither with 9 dither positions to optimize image quality and efficiently remove cosmic rays and bad pixels, and the MEDIUM8 readout pattern to minimize detetor read noise. Given the moderately low galactic latitude of the field at $-$23.7\arcdeg\ it does contain a number of bright Milky Way stars. 

\subsubsection{MIRI imaging}
We used MIRI to image the cluster in 4 filters (Table \ref{smacs_details}). We prioritize depth over coverage, and use a single MIRI tile centered on the cluster, reaching 5$\sigma$ depths of $AB\sim 26.3$\,mag at F770W, $\sim 25.1$\,mag at F1000W, $\sim 24$\,mag at F1500W, and $\sim 23$\,mag at F1800W. We use relatively long ramps with FASTR1 of 100 groups to minimize read noise, except for F1800W, which uses a 50-group ramp given the higher background in this waveband. All images use a 10 point medium cycling dither. 

\subsubsection{NIRISS wide-field slitless spectroscopy} 
We obtained slitless grism spectroscopy with NIRISS of a single tile centered on the cluster, using two different filters to increase the chances of identifying some galaxies that show interesting emission features at high redshift. The F115W and F200W filters are able to detect emission line galaxies via H$\alpha$ at $z=1.0-2.4$ and H$\beta$ along with the [OIII] $4959+5007$\,{\AA} doublet at $z\sim 1.1-3.6$. We used both orthogonal gratings to minimize effects of contamination in the crowded cluster field. Using an 8-point medium dither pattern, and a total exposure time of 2.8\,ks, we reach SNR$\sim$ 10-30 per channel on the continuum for an $AB\sim23\,$mag galaxy, but are also able to detect much fainter emission-line galaxies.

\subsubsection{NIRSpec}
The NIRSpec multi-shutter array (MSA) was used in its medium-resolving configuration to investigate selected galaxies in greater detail, and with wider spectral coverage, reaching 5\,$\mu$m. The medium resolution grating balances sensitivity, velocity resolution, and the number of spectra that can be fitted on the NIRSpec detectors. The G235M and G395M gratings include H$\alpha$ at $z=1.7-6.9$, and the easily identifiable trio of the H$\beta$ line and [OIII] $4959+5007$\,{\AA}    doublet out to $z\sim 9$. We use a 3 shutter slitlet with matching 3-dither pattern, using the NRSIRS2 readout pattern to limit 1/f noise. 

The NIRCam 6-band images, obtained as part of this program, were to derive a catalog and photometric redshifts using the Eazy package \citep{Brammer08}. We used an astrometric solution referenced to Gaia, and supplemented by the RELICS HST/ACS imaging in the area outside the NIRCam footprint \citep{Coe19}. We select galaxies in the field out to $z\sim10$ with $AB<26\,$mag. For the handful of $z>6$ galaxy candidates we included objects with $AB\sim27\,$mag, anticipating that strong emission lines will still easily be detectable. Empty shutters were included on the configuration for subtracting the background. 

Among a total set of 41 primary and 4 secondary targets, we included the 4 highest redshift objects at $z \sim 6-10$, the giant gravitationally lensed arcs, galaxies at $z \sim 1-6$, and cluster galaxies at $z \sim 0.4$. 

\subsection{Stephan's Quintet}
\begin{figure*}[ht!]
    \centering
    \includegraphics[width=18cm]{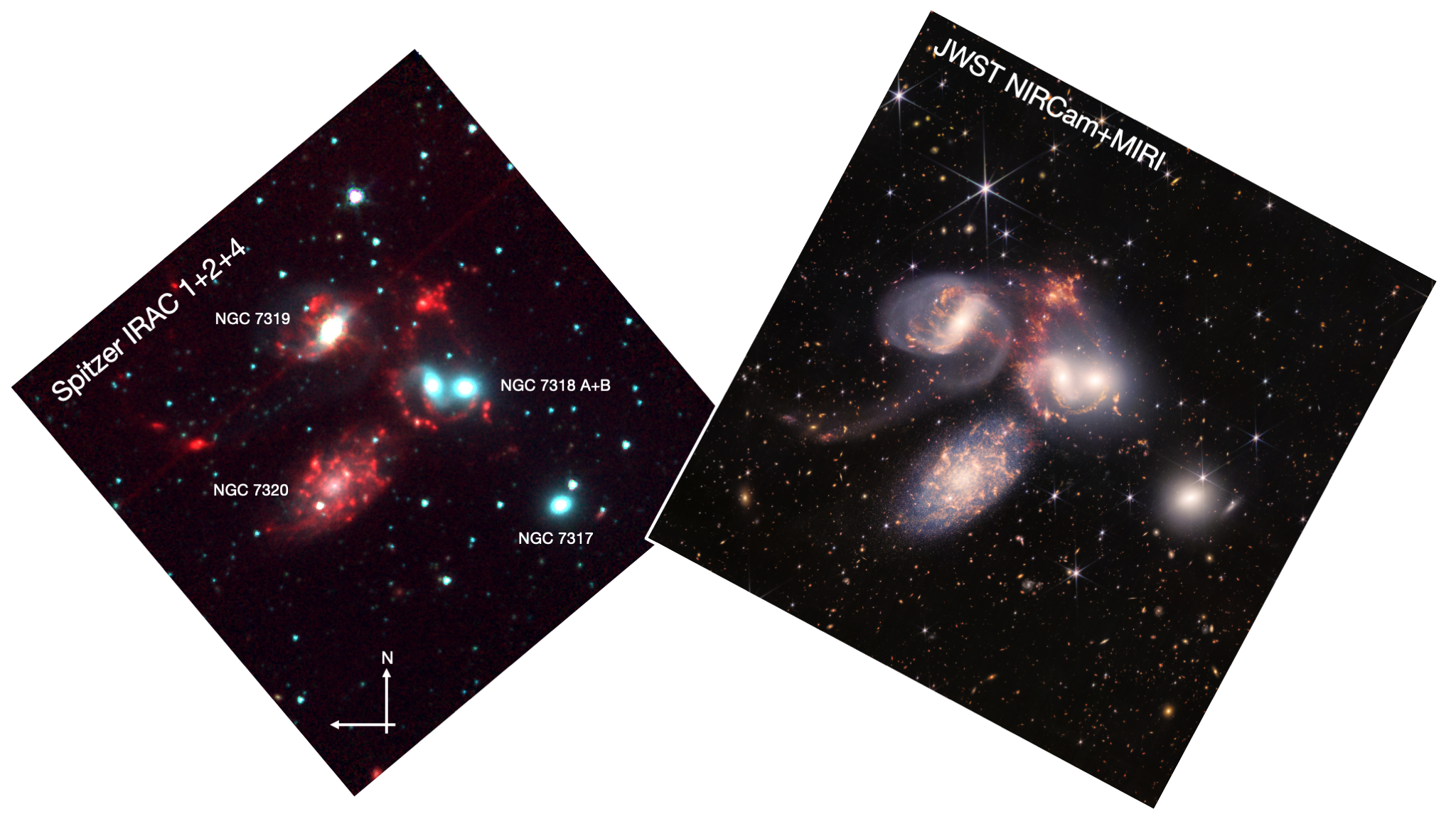}
    \caption{Field of view in Stephan's Quintet covered by NIRCam and MIRI. Full-resolution JWST images are available on webbtelescope.org.}
    \label{fig:stephan_field}
\end{figure*}

Stephan's Quintet is a well-known Hickson Compact Group (HCG), consisting of at least 4 individual large galaxies at an average distance of 88.6 Mpc \citep{DuartePuertas19}, which are likely actively merging. One of the merging galaxies, NGC7319, harbors a bright Seyfert 2 active galaxtic nucleus associated with a radio jet \citep{Aoki99}. A fifth galaxy, NGC7320, is not interacting with the group, as it is much closer \citep{Burbidge61}, and is an apparent chance alignment (Fig.~\ref{fig:stephan_field}).

In the mid-infrared, Stephan's Quintet is characterized by a large-scale shock at the interface between the NGC 7318 and NGC 7319 galaxies, with strong, extended emission from rotational H$_2$ lines \citep{Cluver10}, matching similarly strong X-ray emission \citep{O'Sullivan09}. The objective of the ERO was to illustrate the energetic interactions in the compact group, and show the active growth of an AGN in NGC 7319, all within the context of a field of distant background galaxies.

\begin{deluxetable*}{ccccc}[ht!]
\tablecaption{Observation parameters for Stephan's Quintet / PID 2732}
\label{stephan_details}
\tablehead{
\colhead{Instrument} & \colhead{Mode} & \colhead{Filter/Disperser} & \colhead{Exposure Time$^a$} & \colhead{Observing date}\\
&&& Seconds per Filter/Disperser &
} 
\startdata
NIRCam & Imaging & F090W/F150W/F200W/F277W/F356W/F444W & 2,362 & 2022 Jun 11\\
MIRI & Imaging & F770W/F1000W & 1,332 & 2022 Jun 12, Jul 1\\
MIRI & Imaging & F1500W & 1,354 & 2022 Jun 12, Jul 1\\
NIRSpec & IFU & Prism & 1,284 & 2022 Jun 15\\
MIRI & MRS & SHORT/MEDIUM/LONG & 899 & 2022 Jun 20 \\
\enddata
\tablecomments{$^a$Maximum depth in field}
\end{deluxetable*}

\subsubsection{NIRCam Imaging}
The NIRCam image of Stephan's Quintet is the largest mosaic of the ERO release, consisting of 3 pairs of dithered tiles, arranged such that each pair has a 71.5\% overlap in the direction along the two modules. In total, the final product includes 153.5 Mpixels. The double-tile strategy was designed to maximize the uniformity of the depth across the image to support the production of even-looking color images, and to maintain a minimum of 5 dithers on any given location to be able to produce images free of cosmic rays and bad pixels. Further, we used the FULLBOX dither pattern to image a rectangular field, while minimizing the need to crop irregular edges. 

Since the redshift of the merger is large enough that narrow-band filters will generally not contain their target lines, wide filters are used spanning a wide range in wavelengths from 0.9 to 4.5\,$\mu$m (see Table \ref{stephan_details}). The shortest band includes structure from dark dust lanes within each galaxy, while the longest bands enhances the field of background galaxies, and includes emission from the 3.3\,$\mu$m PAH feature and the strong rotational S(9) line from H$_2$. The exposure parameters were chosen to maximize dynamic range (high sensitivity, but without saturating too many brighter sources), including the brightest areas around the NGC7319 AGN core. The NIRCam images reach 5$\sigma$ depths of $AB\gtrsim 29\,$mag in the deepest areas of the field.

\subsubsection{MIRI Imaging}
Since MIRI covers less than a quarter of the field of view of a single NIRCam tile, there was insufficient time to cover the full NIRCam area. Consequently, the MIRI image covers the central galaxies, NGC7318, NGC7319, and NGC7320, using 4 tiles. The 2 tiles centered on NGC7320 were observed significantly later than the tiles on NGC7318 and NGC7319, and consequently the relative aperture position angles differ by a few degrees. The MIRI image consists of 3 filters, F770W, F1000W, and F1500W, intended to trace PAH emission, as well as to reveal embedded star-forming clusters. The pointing including the NGC7319 core was a critical inclusion, needed for context with NIRSpec IFU and MIRI MRS observations of the Seyfert 2 AGN. We used 8 dither points of the large cycling pattern to maximize the spatial coverage of a single tile. 

\subsubsection{NIRSpec Integral Field Spectroscopy}
We obtained a single NIRSpec IFU tile centered on the Seyfert 2 AGN in NGC7319. We use the PRISM for maximal sensitivity and instantaneous spectral range (0.7-5.2\,$\mu$m). It is noteworthy that the redshift of NGC7319, albeit relatively small, allows the rest-frame optical H$\alpha$ line to be included in the NIRSpec coverage. Sensitivity at shorter wavelengths was a particular consideration, given how red (and likely embedded) the AGN is. At the same time, the IFU exposure parameters avoid saturation at the longer wavelengths. 8 dither points from the large cycling pattern maximizes the spatial resolution. The single tile covers the central parts of the AGN outflow/jet structure.

\subsubsection{MIRI Medium Resolution Spectroscopy}
A single MIRI Medium Resolution Spectroscopy (MRS) tile was also observed, centered on the same region the NGC7319 core that was covered by the NIRSpec IFU. We observed all three sub-bands for full spectral coverage from 4.9--28\,$\mu$m. The MIRI MRS observation was supported by a background observation, centered on a blank part of the sky off the galaxy group, to help remove the telescope background. Given the brightness of the AGN, the MIRI spectrum generally has very high signal-to-noise ($>$100) across the covered spectral range. 

\subsection{NGC 3324 (Carina)}
\begin{figure*}[ht!]
    \centering
    \includegraphics[width=18cm]{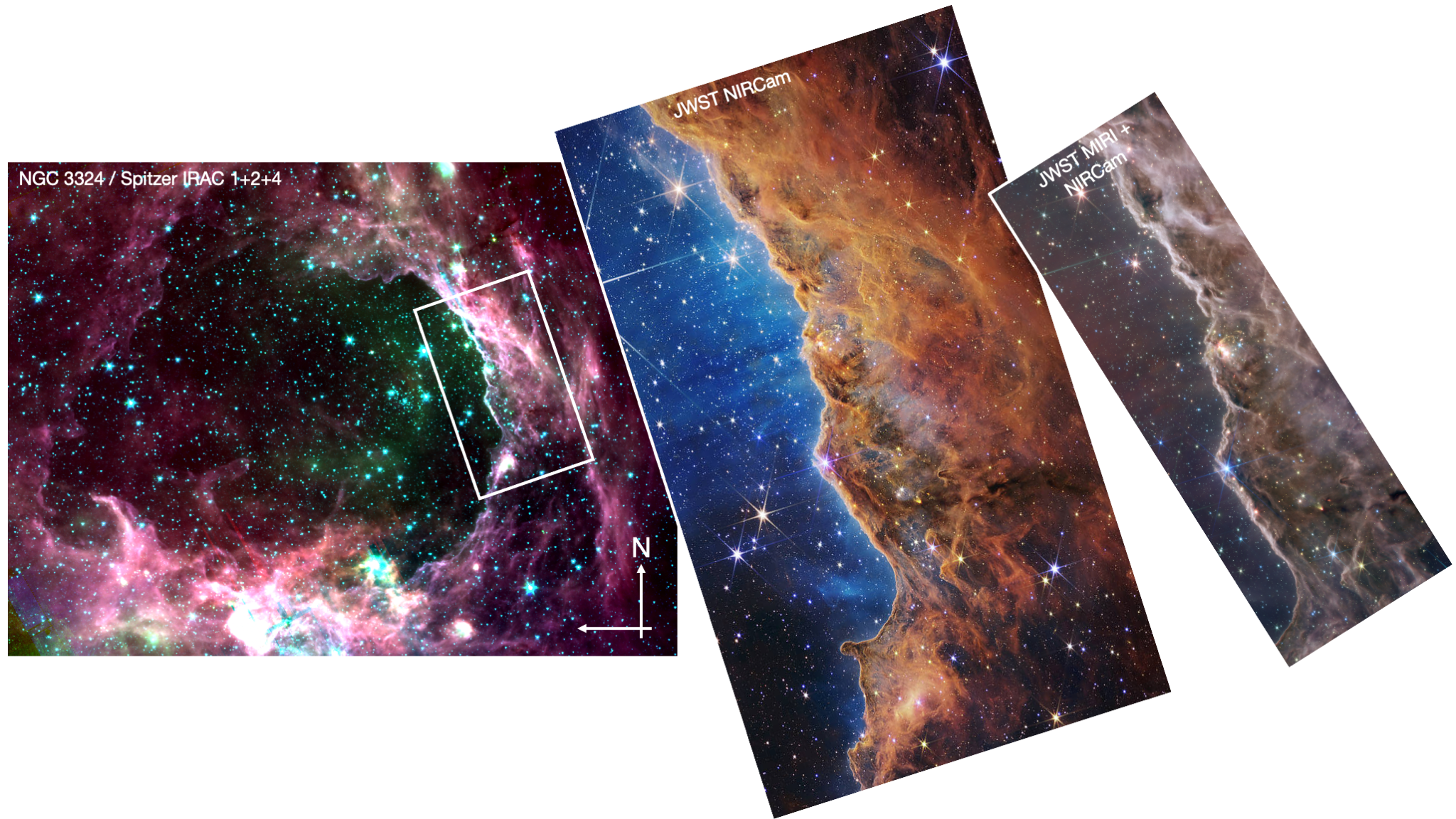}
    \caption{Field of view in NGC 3324 covered by NIRCam and MIRI. The framing was selected using archival images from Spitzer, as shown. Full-resolution JWST images are available on webbtelescope.org.}
    \label{fig:carina_field}
\end{figure*}

The ERO release prominently featured an image of the eastern edge of NGC 3324, an ionized bubble, formed by hot young stars in the north-eastern part of the Carina star-forming complex. NGC 3324 is located at a distance of about 2.2\,kpc \citep{Goppl22}. Named the ``Cosmic Cliffs'', the image captures a landscape view of the photo-dissociation region of the bubble, and highlights the contrast between ionized and molecular gas (Fig.~\ref{fig:carina_field}). 

The eastern edge was chosen as it was expected to show the sharpest ionization boundary, based on the Spitzer image, and offers the greatest brightness and color contrast. The eastern edge was also favorably aligned with the available aperture position angle of NIRCam at the time of observation. The pointing was slightly shifted east to feature more of the molecular nebulosity, which was expected to exhibit the most detailed structure in the infrared. The NGC 3324 field includes a prominent pillar in the south, and is known to exhibit significant star formation, as evidenced by the presence of multiple protostellar jets \citep{Smith10, Ohlendorf13}.

\subsubsection{NIRCam imaging} 
A key objective of the Carina image was to demonstrate, not only how the infrared range can be used to reveal embedded young stars, but also how wavelengths beyond $\sim$2\,$\mu$m trace bright molecular emission from particularly PAHs and H$_2$, leading to spectacular, high-resolution vistas.  

To this end, we observed 4 mosaic tiles with NIRCam, using a pairwise overlap of 71.5\% in the horizontal direction in order to get as much uniformity in the depth as possible. This is the same strategy as that employed for Stephan's Quintet. 

We used the FULLBOX 5-point dither pattern (10 dither points with the overlapping tiles) to be able to construct a rectangular image. For use with color images of galactic star-forming regions, we designed the following 6-filter set: F090W and F200W to image scattered light from dust and show extinction colors of the background stellar field, while optimizing spatial resolution and sensitivity at the shortest wavelengths. The F187N filter captures ionized gas via the bright Pa$\alpha$, F470N images H$_2$ from embedded jets and outflows, F335M images emission from the 3.3\,$\mu$m PAH band, and F444W provides a long-wave sensitive point to dust scattering from large grains. Exposure times are set to not saturate on extended nebular emission, although many point sources will inevitably saturate in the broad filters (see Table~\ref{carina_details}). In particular the F335M filter is critical in bringing out highly textured structure from the surface of the molecular cloud; the PAH emission traces a combination of cloud density and the strength of the local ultra-violet field. Together, these aspects creates a thin layer tracing a surface through the molecular cloud. Since the 3.3\,$\mu$m feature is the shortest wavelength PAH feature, this filter provides the highest resolution view of the molecular boundary layer of the photo-dissociation region available to JWST. 

\subsubsection{MIRI imaging}
As in the case of Stephan's Quintet, the field is too large to completely cover with MIRI imaging. However, the MIRI image mosaic does use 5 tiles to cover a significant fraction of the interface region of the Carina bubble, overlapping with the NIRCam field of view, and including several protostellar associations and associated outflows. An 8 point cycling large dither pattern was used to provide good depth coverage across the panorama. We used 4 filters relevant to star forming regions: F770W and F1130W for PAH emission, F1280W to include emission from the 12.81\,$\mu$m [NeII] line, and F1800W for a long-wavelength point tracing cooler dust and the presence of thick protoplanetary disks around the young stars in the field. Exposure times are set to not saturate the nebular emission, but some bright point sources inevitably saturate (see Table~\ref{carina_details}). 

\begin{deluxetable*}{ccccc}[ht!]
\tablecaption{Observation parameters for NGC 3324 / PID 2731 (Carina)}
\label{carina_details}
\tablehead{
\colhead{Instrument} & \colhead{Mode} & \colhead{Filter/Disperser} & \colhead{Exposure Time$^a$} & \colhead{Observing date}\\
&&& Seconds per Filter/Disperser &
} 
\startdata
NIRCam & Imaging & F090W/F200W/F335M/F444W/F356W & 1,610 & 2022 Jun 3\\
NIRCam & Imaging & F187N/F470N & 2,899 & 2022 Jun 3\\
MIRI & Imaging & F770W/F1130W & 1,354 & 2022 Jun 11\\
MIRI & Imaging & F1280W & 1,399 & 2022 Jun 11\\
MIRI & Imaging & F1800W & 1,199 & 2022 Jun 11\\
\enddata
\tablecomments{$^a$Maximum depth in field}
\end{deluxetable*}

\subsection{NGC 3132 (Southern Ring Nebula)}

As a demonstration of JWST's ability to better understand the full lifecycle of stars, the young planetary nebula NGC 3132 (the ``Southern Ring'') was imaged by NIRCam and MIRI (see Table~\ref{ring_details}). Located at a distance of $\sim$760\,pc \citep{Chornay21}, NGC 3132 contains a visual binary star, one being a bright main-sequence A star, the other a young white dwarf -- the origin of the nebula and also the ionizing source \citep{Ciardullo99}. In the near-infrared, the NGC3132 nebula is characterized by hydrogen recombination lines, lines from highly ionized atomic species ([Ar III], [S IV], [Ne II], and [S III]), and H$_2$ rovibrational and rotational lines \citep{Mata16,Monreal-Ibero20}. The intrinsic morphology of the nebula is likely bipolar, but viewed close to pole-on \citep{Monteiro00}.

\begin{deluxetable*}{ccccc}
\tablecaption{Observation parameters for NGC 3132 / PID 2733 (Southern Ring Nebula)}
\label{ring_details}
\tablehead{
\colhead{Instrument} & \colhead{Mode} & \colhead{Filter/Disperser} & \colhead{Exposure Time$^a$} & \colhead{Observing date} \\
&&& Seconds per Filter/Disperser 
} 
\startdata
NIRCam & Imaging & F187N/F212N/F405N/F470N & 2,319 & 2022 Jun 3 \\
NIRCam & Imaging & F090W/F356W & 1,460 & 2022 Jun 3 \\
MIRI & Imaging & F770W/F1130W/F1280/F1800W & 1,354 & 2022 Jun 12 \\
\enddata
\tablecomments{$^a$Maximum depth in field}
\end{deluxetable*}

\subsubsection{NIRCam imaging}
\begin{figure*}[ht!]
    \centering
    \includegraphics[width=18cm]{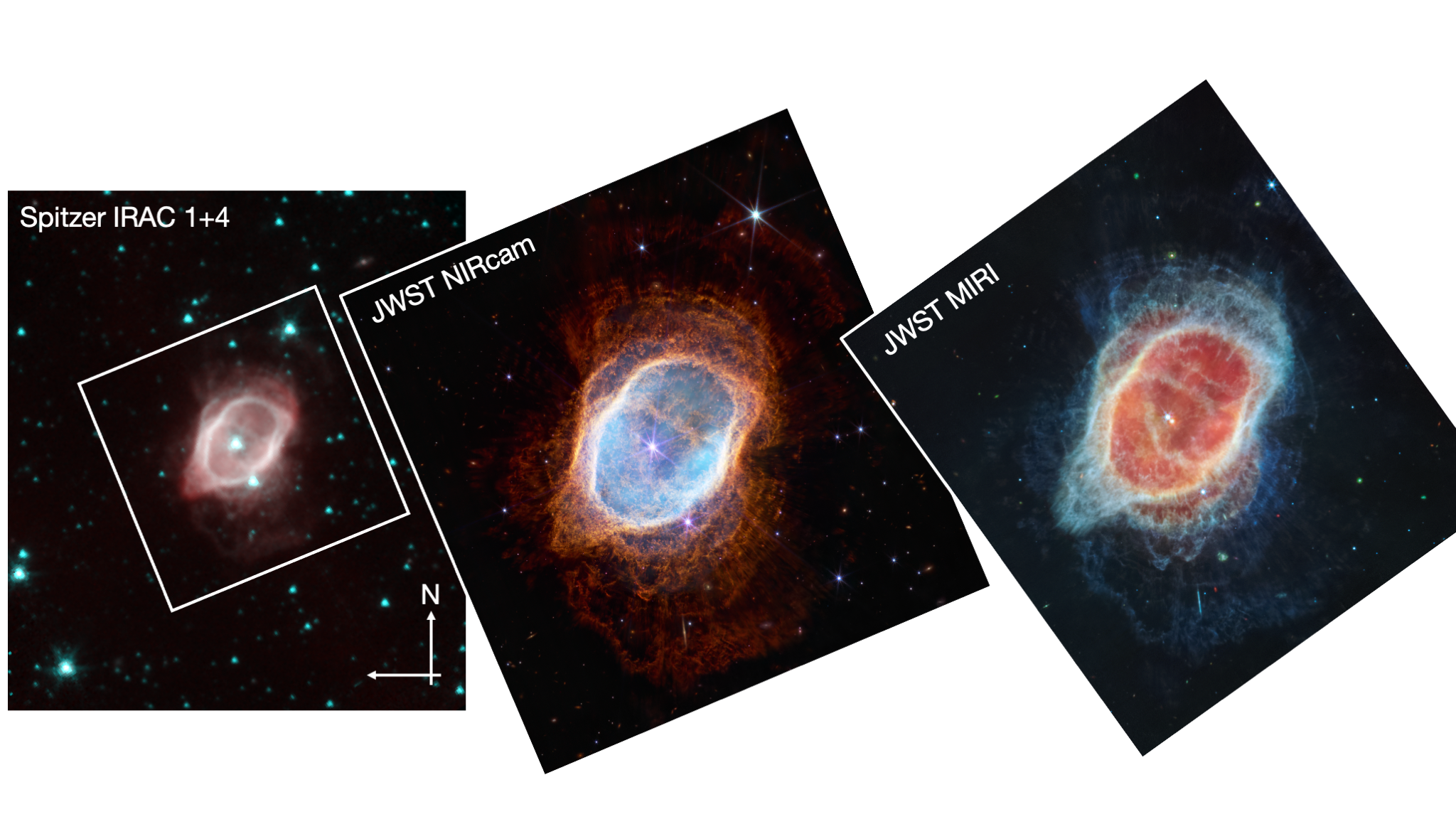}
    \caption{Field of view of NGC 3132 covered by NIRCam and MIRI. The framing was selected using archival images from Spitzer, as shown. Full-resolution JWST images are available on webbtelescope.org.}
    \label{fig:southern_field}
\end{figure*}

We imaged the nebula with a combination of broad-band and narrow-band filters, targeting known spectral features: F090W for the background galactic field and scattered light on dust. F356W was used to detect any PAH emission and to add colors of background galaxies in combination with F090W. F187N and F405N trace the bright Pa$\alpha$ and Br$\alpha$ hydrogen recombination lines, respectively. Finally, F212N and F470N image the strong rovibrational and rotational H$_2$ emission lines in the nebula, tracing different excitation temperatures. The narrow-band filter exposure times are set to reach high signal-to-noise while not saturating, to increase dynamic range, and to reveal turbulent, small-scale structure particularly in the H$_2$ emission from the outer parts of the nebula. The exposure time estimates were informed by the surface brightness as measured in archival Spitzer IRAC1+2 images (Figure \ref{fig:southern_field}). The nebula is fully contained within NIRCam module B, and we use the INTRAMODULEX dither pattern with 8 pointings to get uniform coverage. 

\subsubsection{MIRI imaging}
We imaged with MIRI NGC 3132 using 2 tiles to cover roughly the same field of view as NIRCam Module B. The exposure times are set using archival Spitzer IRAC and MIPS images, along with Spitzer spectroscopy to model performance, as the mid-infrared nebula is entirely dominated by emission lines. We target the 11.3 micron PAH specifically, as it is seen in the Spitzer spectrum. We also target F1280W to measure the known bright [NeII] line, and the F1800W filter to image the 18.71\,$\mu$m [SIII] line, as well as any cool dust emission. We use the 8-point cycling dither pattern, and 2 mosaic tiles to roughly cover the NIRCam module B field, albeit at a slightly different position angle.

\subsection{WASP96b}
We obtained observations with the NIRISS Single-Object Slitless Spectroscopy (SOSS) mode of the star WASP-96 in order to monitor a transit event of the exoplanet WASP-96b \citep{w96} on June 21, 2022. The objective of these observations was to obtain a transmission spectrum of the planet, targeting in particular the known water feature observed at 1.4 $\mu$m by \emph{HST} \citep{yip:2021, nikolov:2022}. The observation, which started at 04:06 UT and ended at 10:30 UT, consisted of a 280-integration, 14 groups per integration, exposure. The number of groups in the integration were selected such that the maximum number of counts achieved on each integration was conservatively at about 50\% the saturation level of the NIRISS detector; these were optimized using PandExo \citep{PandExo}. To schedule the event, we used the period and time-of-transit center obtained by \cite{PE2022}, as well as the ExoCTK tools to define the optimal position angles and phase-constraints to capture it \citep{ExoCTK}. 

\section{Data processing}

\subsection{NIRCam imaging}

All the NIRCam imaging observations were processed with the publicly available `jwst' pipeline\footnote{\url{https://github.com/spacetelescope/jwst}}, using the most up-to-date development version available at the time of the ERO observations (version 1.5.2, corresponding to the Final Release Candidate for JWST Build 8.0 delivered to the operational pipeline, with calibration reference files specified by CRDS context 0850). A total of 1,890 NIRCam image files had to be calibrated for the 4 released ERO targets that were observed with NIRCam, each of which had been imaged in 6 filters using multiple exposures, with all 10 NIRCam detectors in most cases. The pipeline consists of 3 stages \citep{JDox}, which were generally run with default parameters, starting from the raw uncalibrated files, through to fully calibrated exposures, and ending with full combined mosaics. A few of the pipeline steps were customized to improve the quality of the resulting products, as described here.

\subsubsection{NIRCam 1/f noise}
One of the characteristics of the NIRCam detectors is the presence of low-level electronic noise that is highly correlated in time, across output amplifiers and reference pixels, referred to as `1/f noise' \citep{Moseley10}. This is manifested as linear stripes, which change gradually in their amplitude across the detector. This electronic effect is at a relatively low level and did not significantly impact the NIRCam images of NGC 3324, NGC 3132, or the bright galaxies in Stephan's Quintet. However, it was noticeable in fainter regions of the images, and especially so for the deep imaging of SMACS J0723.3-7327. The version of the pipeline available at the time of the ERO observations did not include a correction for this, so a simple correction was applied to each exposure. Specifically, we measured the median value across each row, after masking bright sources, and then adjusted the values in each row accordingly. While some additional higher-order variations might still be present, this procedure provided a first-order correction to improve the low-level background quality and noise properties of the resulting mosaics.

\subsubsection{NIRCam Residual Background Correction}
Another aspect of the NIRCam detectors is the presence of an electronic bias level that is removed during calibration by applying a `superbias' reference file. While this process is generally sufficient to remove this electronic bias, small residual offsets in the background level may remain after calibration, and these may be different for each detector. The JWST pipeline includes a `skymatch' step to match and remove any such residual background, and there were no visible residuals remaining for the ERO observations of NGC 3324, NGC 3132, or Stephan's Quintet. However, for the deep image of SMACS J0723.3-7327, some low-level residual background differences remained, visible as sky level offsets between different detectors. Consequently, an additional iteration of background subtraction was carried out by masking bright sources and extending those masks to also exclude fainter emission before re-measuring the background levels. This successfully corrected the remaining residual background differences in these images.

\begin{figure*}[ht!]
    \centering
    \includegraphics[width=18cm]{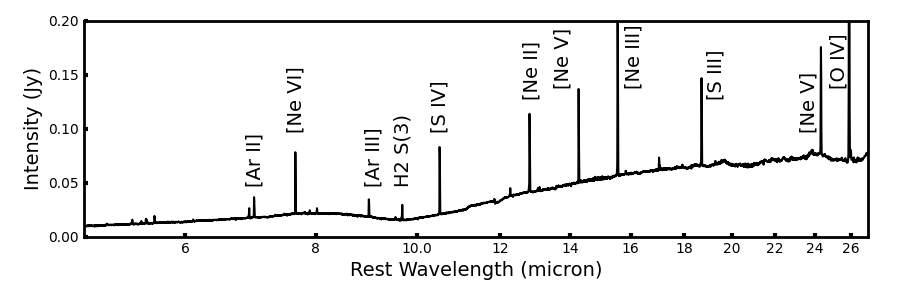}
    \caption{MIRI MRS spectrum of the central 0.65'' radius region surrounding the central Seyfert nucleus of NGC 7319.  A selection of key atomic and molecular emission features have been labelled.}
    \label{fig:ngc7319}
\end{figure*}

\subsubsection{NIRCam Astrometric Alignment}
The NIRCam instrument is composed of two independent modules (A and B), each of which has 5 separate detectors, 4 of which are in the Short Wavelength Channel in a 2$\times$2 grid, with the remaining detector being in the Long Wavelength Channel. For all 10 of these detectors, measurements of their relative positions were available from ground-test data. While preliminary on-orbit measurements had been obtained to refine the detector astrometry, not all of these were available at the time of the ERO observations. Initial calibrations revealed inaccuracies up to $\sim\,$0.2\arcsec\ in the relative locations of some of the detectors, as well as about $+/-$0.5\arcdeg\ in their relative orientations, compared to ground-test data. Therefore, for all the NIRCam images, these offsets had to be corrected by invoking a part of the pipeline that solves for offsets (`tweakreg') in a custom setting, essentially registering each of the 10 detectors independently to an external catalog of the field, solving for their shifts and rotations separately for each filter. For the NIRCam imaging of NGC 3324, NGC 3132, and Stephan's Quintet, the external catalog consisted of Gaia-DR3\footnote{\url{https://www.cosmos.esa.int/web/gaia/dr3}}. For the images of SMACS J0723.3-7327, too few Gaia stars were available to provide a robust direct alignment, so the observations had to be aligned to RELICS HST observations of the field, which also needed to be aligned to Gaia-DR3. This successfully corrected all the astrometric errors, achieving a relative precision of $\sim\,$5 mas and absolute astrometric accuracy of $\sim\,$10 mas, and enabled all the images to be precisely aligned and combined into final mosaics.

\subsection{MIRI imaging}
The MIRI imaging observations were processed with the `jwst' pipeline using the development version (approximately 1.6.0) and with the most up-to-date flight reference files (equivalent to CRDS context 0927). One exception was for the F1800W images, which used a custom flat field created using PID 1052 data, resulting in cleaner mosaics. A small number of extra processing steps were generally applied to improve the quality of the final mosaics, including an additional tweak to improve the astrometry (see Section \ref{sec:miri_astronomy}). Additionally, for the Stephan's Quintet and SMACS J0723.3-7327, additional steps were applied to subtract the average instrumental background and to correct column/row pull up/down around very bright sources.

\subsubsection{Refining Astrometry}
\label{sec:miri_astronomy}
The relative astrometry between dithered images taken using the same guide star is generally accurate to a few mas \citep{Rigby22} -- more than sufficient for MIRI imaging with a pitch of 110~mas.  However, the relative astrometry between mosaic tiles observed using different guide stars can have much larger offsets, driven by the precision of the guide star catalog. Generally, offsets between different guide stars of a few 0\farcs1 are common, with occasional differences as high as 1--2$\arcsec$. While this is enough to cause noticeable registration problems with MIRI imaging mosaics, offsets are usually straightforward to measure and correct for using the `tweakreg' step in the calwebb\_image3 pipeline step, provided that there are a sufficient number of point sources within each tile. This is not always the case for MIRI imaging, especially at longer MIRI wavelengths where stars become very faint.

We refined the MIRI imaging astrometry and registration by deriving offsets between dithers and tiles using `tweakreg' on the shortest wavelength band (e.g., F770W), and then applying the same offsets to every other longer-wavelength band. Further, the offsets in (V2,V3) reported by `tweakreg' for each F770W exposure were averaged over all the dithers within a mosaic tile for enhanced precision and robustness against a lack of point sources sources in each individual dither. This approach takes advantage of the excellent pointing stability and accuracy of the JWST platform. Finally, the `tweakreg' step registers the absolute astrometry to the Gaia frame of reference.

\subsubsection{Instrumental background subtraction}
For observations of faint, low-background targets, residual instrumental structure becomes significant. This was corrected by subtracting an average background image constructed by sigma clipping the stack of all exposures in detector coordinates to remove sidereal sources. While this step significantly reduced the residual instrumental background structure, it did imprint weak, large spatial scale structures due to imperfectly removed astrophysical extended emission.

\subsubsection{Column/row pull up/down correction}
For sources that are very bright compared to the background emission, detector artifacts causing bars to appear along detector columns and rows appear. We used a simple correction by constructing column or row median profiles, and subtracting them from the appropriate detector regions in each exposure.  The sources and regions suffering form these effects were identified visually in the mosaics.  The column/row profiles were constructed and applied independently above/below and left/right of bright sources.

\subsection{NIRSpec Integral Field Spectroscopy}
The JWST NIRSpec integral field unit (IFU) provides spatially resolved imaging spectroscopy over a 3\arcsec\ × 3\arcsec\ square region, and enables spectral imaging over small fields of view. 
The NIRSpec IFU observations of  the AGN in NGC 7319 were processed with the publicly available `jwst' pipeline using  the most up-to-date version available at the time of the
ERO observations (version 1.5.2). 

The observations consist of 8 dither points from the large cycling pattern, which maximizes the spatial resolution. The PRISM/CLEAR combination  was used which produces spectra at a nominal resolving power of R$\sim100$  in the wavelength range 0.6 -- 5.3 microns. 

For the ERO release, we used pre-flight reference files, since in-flight files were not yet available. We started from the output of the first stage of the `jwst' pipeline (calwebb\_detector1) which applies basic detector-level corrections. We ran the rate files through the second stage (calwebb\_spec2), and the calibrated files through the third stage (calwebb\_spec3). In particular, in the cube\_build step of stage 3, we set the weighting mode to ``drizzle'', which is a 3-D generalization of the classical 2-D drizzle technique.

We removed several bad pixels with negative flux values in the final data cube, which we were able to remove by setting the DQ file extension to ``do not use''. Further, the final combined cube had several bad pixels with positive flux values, as well as an edge effect. These two issues were corrected by turning on the outlier\_detection step and setting the scaling factor to ``scale''. A future science-ready data set will require in-flight reference files for spectro-photometric calibration.

\subsection{MIRI Medium Resolution Spectroscopy}

MIRI MRS observations were processed through version 1.6.2 of the jwst pipeline using standard MRS pipeline notebooks similar to those available online\footnote{https://github.com/STScI-MIRI/MRS-ExampleNB} augmented by custom processing scripts. This version of the `jwst' pipeline implements multiple bug fixes compared to previous versions, including a fix to construction of the spectral background model and the residual fringe correction. These notebooks followed the standard pipeline workflow, processing both science and dedicated background observations through the Detector1 and Spec2 pipelines individually before combining the data in the Spec3 pipeline.

First, we made a few additional corrections to the default pipeline workflows in order to improve the quality of the final data products. First, we used manually measured locations of Gaia-DR3 catalog stars in the MIRI simultaneous imaging field to refine the spacecraft pointing information for the associated MRS exposures (which can be uncertain due to a combination of guide star catalog errors and roll uncertainty). This resulted in a correction of about 0.25'' to the astrometric solution of the MRS data.

Second, since contemporal background and science exposures share many of the same detector artifacts, we used the two-dimensional rate files from the dedicated background field to improve the data products from the pipeline Detector1 stage before passing these products into the Spec2 pipeline. This step included ad-hoc column median corrections for vertical striping (due to residuals in the MIRI DARK and RESET steps), median combination of the rate images from individual integrations, instead of the usual pipeline mean combination (to mitigate the effects of large cosmic ray showers), and refinement to the bad pixel flagging by identifying hot pixels in the empty background field.

After passing the rate images through the Spec2 pipeline, we ran the 2-dimensional residual fringe correction step\footnote{\url{https://jwst-pipeline.readthedocs.io/en/latest/jwst/residual\_fringe/index.html}} on the resulting outputs. This step can dramatically decrease the amount of fringing seen in compact point sources coming out of the default pipeline, especially at wavelengths around 11 $\mu$m. Although the necessary code for this correction is part of the JWST pipeline, it is not yet fully integrated into the pipeline workflow and thus must be run by hand. The resulting images were reprocessed to create per-exposure data cubes and one-dimensional spectra.

\begin{figure}
\centering
\includegraphics[width=9cm]{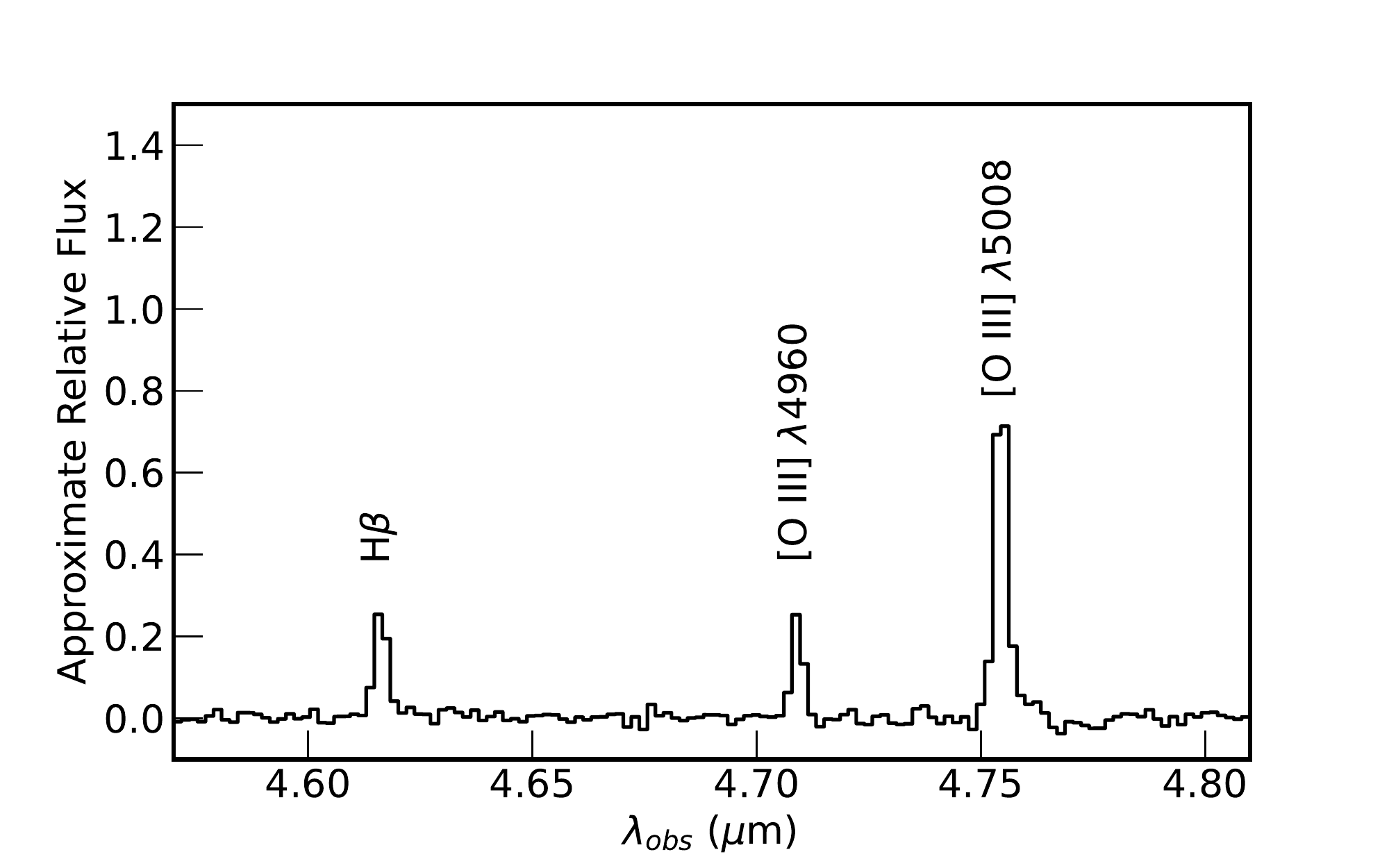} 
\caption{The NIRSpec MOS spectrum of a galaxy at $z = 8.49$ shows robust detections of H$\beta$ and the [\ion{O}{3}] $\lambda \lambda$4960, 5008 doublet. The SNR on the lines ranges from 13 to 44. }
\label{fig_z8}
\end{figure}

Finally, the residual-fringe corrected files were passed into the standard Spec3 pipeline, which combined all of the one-dimensional spectra of the background region to create a model spectrum to subtract from the science data. The final data cubes were constructed using a 3D variant (Law et al. 2022, in prep) of the drizzle algorithm. We opted to configure the final cube building step to produce a single 5-28 micron data cube to ease visual inspection of the results, although we note that such a cube will contain artifacts at long wavelengths due to the sampling characteristics of the cube being optimized for the shortest-wavelength data. Example one-dimensional spectra (e.g., Figure \ref{fig:ngc7319}) were extracted from this data cube using simplified circular apertures; while such apertures do not account for wavelength-dependent flux losses due to the growing PSF, such correction factors are ill-defined for the spatially extended sources with complex morphology in the NGC 7319 field.

\subsection{NIRSpec Multi-Object Spectroscopy}
The NIRSpec MOS observations were processed through version 1.5.2 of the JWST Pipeline. While some in-flight reference files were used, many critical calibrations were not yet available\footnote{The CRDS context describing the NIRSpec MOS reference files was jwst\_0913.pmap.}. As a result, the data were not spectro-photometrically calibrated for use in the press release of the EROs. 

The data were reprocessed through the Spec2 pipeline using a preliminary in-flight S-flat (Ferruit 2022, private communication). Since our emphasis was on the compact sources, we were able to use in-slitlet nodding to subtract the background. In detail, for each nod in the three shutter slitlet, the JWST pipeline averages the two "off" nods and subtracts this 2D spectrum from that of the "on" nod. Finally, the Spec3 pipeline was run, combining the six exposures (3 nods $\times$ two observations) for each of the two gratings.  

After examining the pipeline outputs, we concluded that the default 1D extractions were insufficient, with apertures clearly including the ``negative flux'' from the nod-subtracted spectra. Therefore, we applied a by-hand aperture extraction to the s2d spectra for the selected objects that were under consideration for the ERO release. Finally, additional masking was carried out, thereby removing artifacts that were not flagged by the pipeline. 

Future reprocessing will be required for these data to be science-ready. Specifically, in-flight reference files will be needed to obtain a proper calibration, and further investigation of the pipeline's 1D extraction and data quality flagging will also be necessary. Finally, the construction and application of a master background spectrum from the background shutters remains to be tested. Nonetheless, the apparent quality of these spectra is outstanding. Figure \ref{fig_z8} shows the emission from H$\beta$ and the [\ion{O}{3}] $\lambda \lambda  4960, 5008$ doublet, redshifted to $z=8.49$. The signal-to-noise ratio in these lines ranges from 13 to 44, suggesting exciting potential for these and other NIRSpec MOS observations of galaxies in the early universe.

\subsection{NIRISS Wide-Field Slitless Spectroscopy}
The NIRISS WFSS images were processed using version 1.5.2 of the JWST pipeline. The direct images for the WFSS observations were processed through the standard imaging pipeline workflow, including the Detector1, Image2, and Image3 pipeline steps. The tweakreg step of the default Image3 configuration was modified to precisely register individual dithers. The outputs from the imaging pipeline for each filter includes a fully calibrated image that combines all the dithers, a source catalog, and a segmentation image from the source identification. The dispersed grism images were processed through the Detector1 step of the pipeline and through some of the initial steps of Spec2 pipeline, which include WCS assign, background subtraction, and flat fielding. The dispersed images were then shifted and stacked using custom scripts to improve the signal-to-noise before spectral extraction. The spectral extractions were done using the pre-flight calibration files to locate the spectral traces for the sources of interest (including lensed arcs in the SMACS J0723 field) and custom codes were used to refine the trace positions and extract the spectrum. The wavelength calibration and spectrophotometry were also done using pre-flight calibrations. The NIRISS WFSS data will have to be reprocessed using the new reference files from commissioning to provide accurate wavelength and flux calibrations for future analyses. 

These first data from NIRISS WFSS already reveal a large number of emission line sources and the combination of H$\beta$ and [OIII] $4959+5007$\,{\AA} doublet are easily identified at different redshifts across the detector field-of-view, clearly demonstrating the power of this mode for emission line galaxy surveys. In many cases the emission lines are spatially-resolved in the grism images which highlights the potential of using dispersed spectra to probe variation of nebular properties within distant galaxies.

\subsection{NIRISS Single-Object Slitless Spectroscopy}

\subsection{Detector-level analysis}
The NIRISS/SOSS images were processed using version 1.5.2 of the JWST pipeline. We worked with the Stage 1 products to produce the transmission spectrum, i.e., with the rates per integration. From there, the 2D spectra were analyzed and extracted using custom 
routines specialized to deal with NIRISS/SOSS data\footnote{\url{https://github.com/nespinoza/transitspectroscopy}}. 

First, we traced the Order 1 and Order 2 spectra using a custom routine that performs a cross-correlation of each column in the image with a sum of two Gaussians separated by 15 pixels. Once the position of the center of those traces was obtained, we fitted those with splines in order to smooth their shapes. Next, we performed 1/f noise corrections by subtracting the median of all integrations from each individual integration. This left us with a frame for each integration containing detector-level effects. With these frames, we obtained the level of 1/f noise by computing the median number of counts within 20 to 35 pixels from the trace. These counts were then substracted to each pixel within a 15-pixel radius from each trace of the original integrations. 

The final step on our detector-level data reduction removed the zodiacal background from each integration. Under the assumption that the zodiacal background stays constant throughout the exposure (an assumption we tested with our data), we scaled the background model provided in the JWST Documentation\citep{JDox} using the median of all the integrations in the exposure. In particular, we used columns 500 through 800 and rows 210 through 250 of both the data and the model, taking the ratio of both to find the scaling factor. We used the second quartile of this ratio to estimate the scaling factor between the data and the model, which gave a ratio of about 0.47. This value was then used to remove the zodiacal background signal from each individual integration. The spectra were extracted using simple box extraction with an aperture radius of 15 pixels.

\subsection{Lightcurve fitting}
The 280 spectra implied 280 fluxes measured in each wavelength bin (2044 for Order 1 and 551 for Order 2), which formed our set of 2591 transit lightcurves to analyze. Each lightcurve was divided by the out-of-transit median flux and then fitted using  \emph{juliet} \citep{juliet}. These were fitted using a \emph{batman} model \citep{batman} with the best-fit orbital parameters obtained from an analysis made on the white-light lightcurves fixed in the wavelength-dependent analysis. The limb-darkening coefficients were fitted at each wavelength using the sampling scheme of \cite{kipping} and a square-root law, but using limb-darkening coefficients obtained through the \emph{limb-darkening} software combined with the SPAM algorithm as priors \citep{EJ:2015}. The priors were set on the transformed coefficients $(q_1, q_2)$, with truncated normal distributions centered on these theoretical predictions, having standard deviations of 0.1 and constrained to be between 0 and 1. In addition to the transit model, a Gaussian Process was used to model any trends and/or instrumental systematics. In particular, a Mat\`ern 3/2 kernel in time was used, which defined two extra parameters for the fit: The amplitude of the Gaussian Process and the time-scale of the process. Additional free parameters were a mean out-of-transit flux, a photometric `jitter' term added in quadrature to the error bars for each wavelength, and the planet-to-star radius ratio. We used the dynamic nested sampling algorithm implemented in the \emph{dynesty} package as described in \cite{dynesty} to sample from the posterior distribution of the parameters. While the posterior distribution for all parameters were obtained for our analysis, the only observable presented in the ERO package was the square of the planet-to-star radius ratio --- i.e., the transit depth as a function of wavelength. While the transit depths were calculated at the resolution level of the instrument, we presented a spectrum binned to $R=100$.

\section{Image Visualization}

The process of producing color images from the JWST data is similar to that of other observatories, such as the Hubble Space Telescope. A key objective was to develop a striking translation of infrared colors to the visible color space, using physical and chemical tracers not available in the Hubble range. While previous infrared telescopes, such as Spitzer, made great strides in this direction, JWST offers many more filters, resulting in a much greater number of potential color combinations. The final color images represent one option out of many possible, for a wide range of different types of object, from the deep universe, where more distant or dusty galaxies naturally appear red using a range of broad-band filters, to a planetary nebula entirely dominated by non-thermal line emission from molecular, atomic, and ionized gas. We found that, for JWST, excellent sources of color contrast and visual impact include 1) the broad emission bands of PAHs, with the 3.3\,$\mu$m PAH feature offering the highest spatial resolution and structure, while the 7.7 and 11.3\,$\mu$m bands providing stunning views of the interstellar medium in nearby galaxies, 2) emission lines from rotational and ro-vibrational transitions from H$_2$, with the S(9) line at 4.7\,$\mu$m providing a good balance between spatial resolution and brightness for shocks and outflows, and 3) ionized gas traced by the 1.87\,$\mu$m Pa$\alpha$ hydrogen recombination line and the 18.7\,$\mu$m [SIII] line as contained within the F1800W MIRI filter. 

\subsection{Scaling}
The dynamic range of the image data was first compressed using FITS Liberator\footnote{\url{https://noirlab.edu/public/products/fitsliberator/}} and PixInsight\footnote{\url{https://pixinsight.com/}}. Typically, an ArcSinH stretch coupled with a high scaled peak value works best to reveal the most low-level signal while preserving the brightest features. The stretch must be applied carefully, however, as not to exacerbate background noise. 

\subsection{Noise Suppression and Artifact Removal}
Once scaled, each filter must be treated individually for artifacts and noise intrinsic to the detector. The most common noise features and artifacts found in the JWST data include saturated star cores, banding from residual 1/f noise, gradients/background variations, and ‘persistence’- typically where a bright star leaves a remnant image in succeeding exposures. 

To recover the saturated cores of the stars, an algorithm, PixelClip.js, adapted by J. DePasquale from G. Barrere (Priv. Comm.), is applied, which replaces the values below a certain threshold with the mean of its nearest neighbor. The original algorithm replaced saturated values above a certain threshold, but this was modified to deal with JWST data where the saturated regions are filled with null values. 

Any residual horizontal and vertical banding from 1/f noise is removed by applying a photoshop action found within the ‘Astronomy Tools’ package, which reduces noise by scanning across the image a couple pixels at a time while attempting to match the sky background. This action can only work if the banding is parallel to detector rows and columns, so the image must be aligned accordingly before running this step. The procedure is also not as effective on larger mosaics such as Stephan’s Quintet and Carina, but whatever banding remained was addressed by creating an artificial background which was used to subtract out the 1/f noise, while preserving the signal. This method was also applicable to dealing with background gradients and other variations along with manual ‘curves’ adjustments.

Persistence is handled in the same way as residual cosmic rays, where the background sky nearest to the artifact is sampled and replaced with that selection.

\subsection{Chromatic Ordering and Aesthetics}
In order to achieve the most scientifically illustrative and aesthetically pleasing visual impact, chromatic ordering is applied almost exclusively. There are a very small number of exceptions where the scientific value gained outweighs a minor break from this convention. For instance, in the Carina NIRCam image, we assigned a yellow, rather than red, color to the long F470N filter tracing H$_2$ emission. However, in this case, this was justified as there are many available H$_2$ lines throughout the infrared range, and the 4.7\,$\mu$m S(9) line is only one option of many. This means, color is applied based off the wavelengths of the filters, where shorter wavelengths get prescribed a bluer color, while longer wavelengths get prescribed redder color. The exact colors that get assigned, depends on the data, and on what combination provides the most detailed and interesting image. We generally considered a range of color balance options, before settling on a final version that offers the best trade-off between science and aesthetics. 

Once color has been assigned and the filters have been combined, some technical and aesthetic adjustments remain. White balancing is critical in getting a natural looking image. One method is to use the star cores as a white reference, which often yields a desirable result or at least a good starting point. A more accurate method, is to use background spiral galaxies, which represent the entire range of the spectrum, as a white reference. It is also important to make sure the background sky is truly neutral in order to achieve a balanced image.

The image visualization is naturally a subjective process, but the goal is always to enhance the astrophysical features and processes in each filter through contrast adjustments, color separation and other techniques while preserving the intrinsic integrity of the data.   

\begin{acknowledgements}
This work is based on observations made with the NASA/ESA/CSA James Webb Space Telescope. The data were obtained from the Mikulski Archive for Space Telescopes at the Space Telescope Science Institute, which is operated by the Association of Universities for Research in Astronomy, Inc., under NASA contract NAS 5-03127 for JWST. These observations are associated with programs \#2731, \#2732, \#2733, \#2734, and \#2736. This work is based in part on observations made with the Spitzer Space Telescope, which was operated by the Jet Propulsion Laboratory, California Institute of Technology under a contract with NASA. We acknowledge the foundational efforts and support from the JWST instruments, STScI planning and scheduling, Data Management teams, and Office of Public Outreach.
\end{acknowledgements}

\bibliography{firstimages}{}
\bibliographystyle{aasjournal}



\end{document}